\documentclass[letterpaper, 10 pt, conference]{ieeeconf}  

\IEEEoverridecommandlockouts                              

\usepackage{graphics} 
\usepackage{epsfig} 
\usepackage{mathptmx} 
\usepackage{times} 
\usepackage{amsmath} 
\usepackage{amssymb}  
\usepackage{subfigure}
\usepackage{float}

\title{\LARGE \bf
	A Novel Method for Extrinsic Calibration of Multiple RGB-D Cameras
	Using Descriptor-Based Patterns
}

\author{Hang Liu$^{1}$, Hengyu Li$^{1*}$, Xiahua Liu$^{2}$, Jun Luo$^{1}$, Shaorong Xie$^{1}$, Yu Sun$^{1,3}$
	\thanks{$^{1}$Hang Liu, Hengyu Li, Jun Luo, Shaorong Xie, and Yu Sun with the School of Mechatronic Engineering and Automation, Shanghai University, China. Hengyu Li is the corresponding author. {\tt\small scholar.hang@gmail.com} {\tt\small lihengyu@shu.edu.cn}}%
	\thanks{$^{2}$2Xiahua Liu with the School of Mechanical Engineering, Beijing Institute
		of Technology.}%
	\thanks{$^{3}$Yu Sun with the Department of Mechanical and Industrial Engineering, University of
		Toronto, Canada, and the Shanghai University. {\tt\small sun@mie.utoronto.ca}}%
}

\begin{document}

	\maketitle
	\thispagestyle{empty}
	\pagestyle{empty}

	\begin{abstract}
		This letter presents a novel method to estimate the relative poses between RGB-D cameras with minimal overlapping fields of view in a panoramic RGB-D camera system. This calibration problem is relevant to applications such as indoor 3D mapping and robot navigation that can benefit from a 360$^\circ$ field of view using RGB-D cameras. The proposed approach relies on descriptor-based patterns to provide well-matched 2D keypoints in the case of a minimal overlapping field of view between cameras. Integrating the matched 2D keypoints with corresponding depth values, a set of 3D matched keypoints are constructed to calibrate multiple RGB-D cameras. Experiments validated the accuracy and efficiency of the proposed calibration approach, both superior to those of existing methods (800 ms vs. 5 seconds; rotation error of 0.56 degrees vs. 1.6 degrees; and translation error of 1.80 cm vs. 2.5 cm.
	\end{abstract}

	\section{INTRODUCTION}
	In recent years, indoor scene reconstruction and robot navigation have attracted much attention with the advent of low-cost and efficient depth and color (RGB-D) devices such as the Microsoft Kinect, Intel RealSense, and Structure Sensor. The depth cameras of these devices can provide a depth map with a VGA resolution (640x480) at video rate (e.g., 30 Hz) using efficient light-coding technologies that avoid the challenging task of dense 3D reconstruction from color images. The 3D models reconstructed using these depth cameras have been used to generate more realistic 3D content for virtual reality (VR) \cite{Salas2013SLAM} and help align the rendered virtual objects with real scenes for augmented reality (AR) \cite{newcombe2011kinectfusion}. Furthermore, the direct depth sensing capability of these depth cameras is particularly suitable for robots to navigate in an unknown environment.
	
	With an RGB-D camera, the simultaneous localization and mapping (SLAM)-based approach is mainly used for fusing point cloud frames to reconstruct indoor scenes \cite{mur-artal2017orb, whelan2016elasticfusion, Endres20173, Yan2017Dense}. However, hundreds or even thousands of frames must be captured in state-of-the-art SLAM systems to reconstruct a common indoor environment, such as a room or an office \cite{sturm12iros} because of two problems. (1) The field of view (FoV) of depth cameras is limited; thus, only a small part of the scene is represented in a single frame. The Kinect, for example, has a horizontal FoV of 57$^\circ$, which is much smaller than the horizontal 240$^\circ$ FoV of the Hokuyo URG-04LX-UG01, a laser scanner with a similar range and measurement accuracy to the Kinect \cite{zug2012laser}. (2) To track the poses of depth cameras to effectively fuse multiple point cloud frames, consecutive frames must be captured to have sufficient scene overlap. Typically, more than ninety percent of overlap is required, which further increases the number of frames for reconstruction.
	
	One solution to these problems is to use a multi-camera setup in which RGB-D cameras face different directions to sample different sections of the environment \cite{Fernandez2014Extrinsic, Rives2016Scene, Gokhool2014A, Gokhool2015A}. However, complications can occur with the use of multiple cameras, such as more difficult calibration caused by the minimal overlapping FoV between cameras.
	
	\begin{figure}[t]
		\centering
		\subfigure[]{\label{fig11} \includegraphics[width=0.15\textwidth]{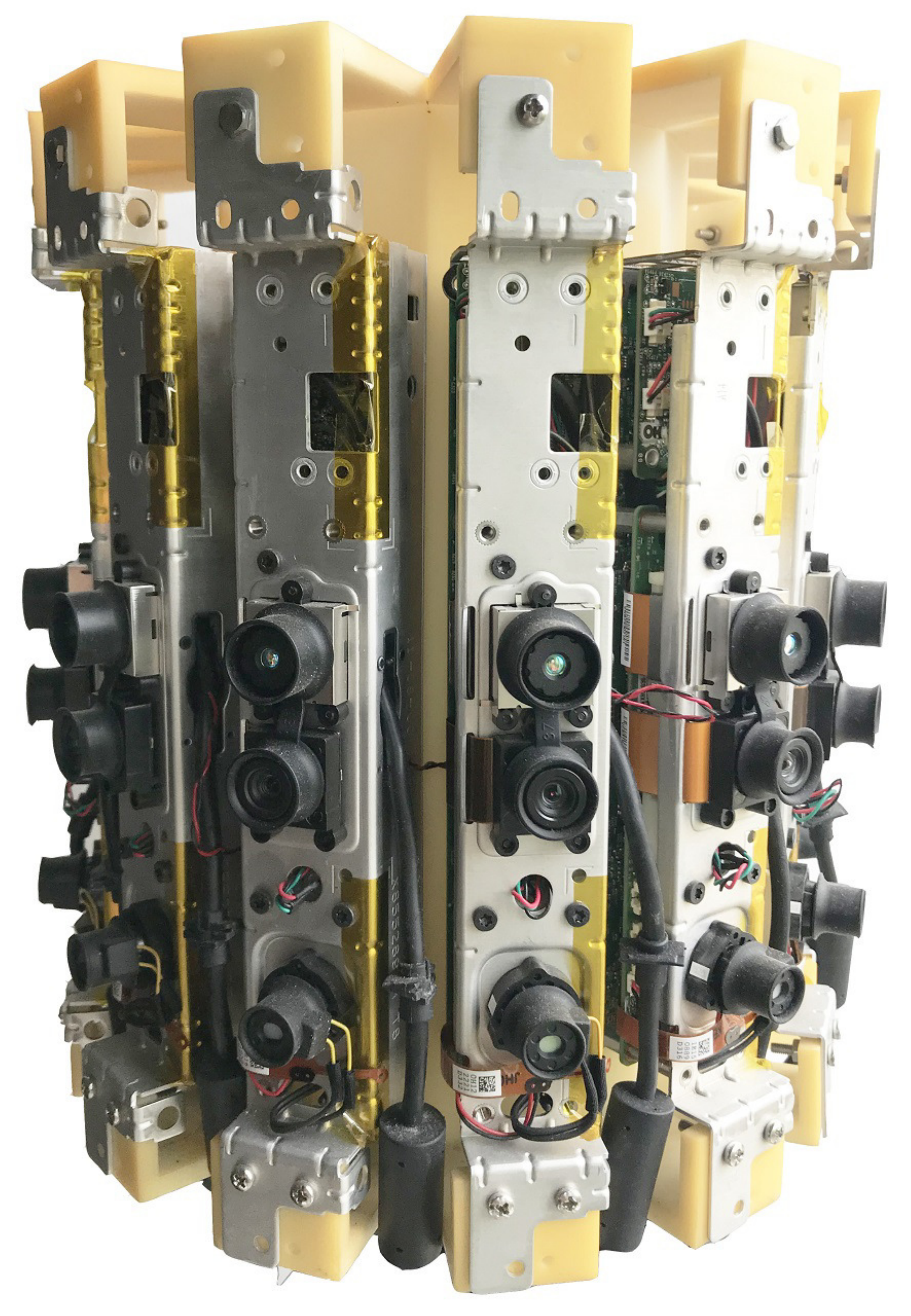}}
		\subfigure[]{\label{fig12} \includegraphics[width=0.315\textwidth]{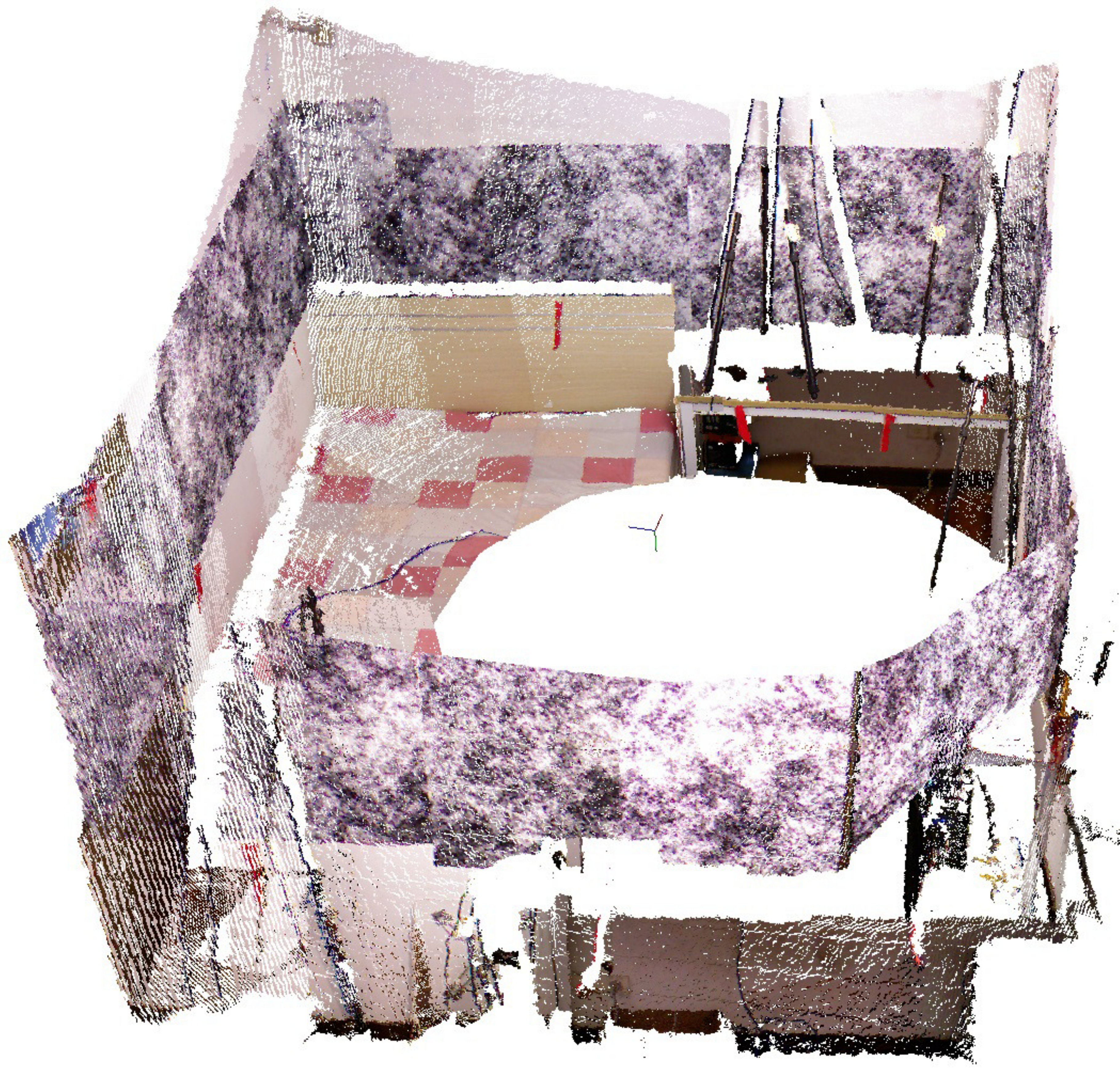}}
		\caption{(a) The setup used in this study for evaluating the proposed calibration method composed of 12 Kinect v1 RGB-D cameras. (b) The panoramic 3D color point cloud obtained by the panoramic 3D vision system. Descriptor-based calibration patterns are pasted on the walls to calibrate multiple RGB-D cameras.}
		\label{Fig1}
	\end{figure}
	
	Classical extrinsic calibration strategies such as the chessboard-based method and the keypoints-based method cannot be applied to calibrate RGB-D cameras in a multi-camera setup, because their overlap requirement constitutes a very strong constraint. A more general approach that can calibrate multiple cameras in an arbitrary configuration is based on per-camera odometry \cite{carrera2011slam,heng2013camodocal,schneider2013odometry,brookshire2012extrinsic}. Cameras are calibrated by finding all camera odometry transforms based on matched features from frames that are captured in the motion paths of all cameras. In these methods, SLAM or visual odometry techniques are applied to estimate camera trajectories. However, the robustness of SLAM and visual odometry techniques highly depends on the environment.
	
	Instead of the tedious estimation of useful trajectories for calibration, Fernandez-Moral et al. proposed to calibrate multiple cameras through planes and lines \cite{Fernandez2014Extrinsic, Perez2017Extrinsic}. Planes and lines have large spatial spans; thus, they can be observed by cameras with little or no overlapping FoV. However, to apply the Fernandez-Moral method, the multi-camera setup is required to be moved around in scenes to extract enough matched planes or lines, which results in low efficiency.
		
	In this work, a new extrinsic calibration method that relies on descriptor-based patterns to provide well-matched 2D keypoints is proposed to estimate the relative poses between the RGB-D cameras with minimal overlapping FoV. In our method, a set of 3D matched keypoints are constructed based on extracted 2D keypoints and corresponding depth values from depth maps to directly estimate poses between multiple RGB-D cameras. Then the estimated poses are globally optimized through the loop-closure constraint provided by the panoramic camera setup (Fig. \ref{fig11}). Experimental results quantitatively verified the accuracy of this method and demonstrated that it is fast and easy to apply.

	\section{EXTRINSIC CALIBRATION}
	
	In this section, we address the problem of estimating extrinsic calibration (i.e., relative poses) between RGB-D cameras that have little overlapping FoV. Fig. \ref{fig11} shows the panoramic RGB-D camera setup that we built to evaluate the proposed calibration method. It consists of 12 Kinect v1 cameras, all Kinects are vertically positioned for a more compact design. The Kinect v1 camera has an angular field of view (FoV) of 43$^\circ$ from the vertical. The overlap FoV of two neighboring cameras is only approximately 30 percent of the vertical FoV of each camera.
	
	We propose to solve the calibration problem of multiple RGB-D camera with little overlapping FoV using feature descriptor-based calibration patterns \cite{Li2014A} (see the patterns on the walls in Fig. \ref{fig12}), which can provide robust and accurate matched feature points in this case of minimal overlapping FoV. Based on these matched 2D feature points and depth maps from the depth cameras, we construct two 3D point sets to estimate poses by bundle adjustment. Then, a pose graph optimization method is used to refine the estimated poses.
	
	\subsection{Initial Estimation of Poses}
	The descriptor-based calibration pattern is composed of several noise images at different scales in accordance with the mechanism of SIFT/SURF. Compared with natural scenes, this pattern contains a high number of detectable features that can be easily detected by a camera at varying distances. Thus, the descriptor-based pattern can provide many and more accurately matched keypoints between two cameras in the case of minimal overlapping FoV.
	
\begin{figure}[t]
		\centering
		\subfigure[]{\label{fig21} \includegraphics[width=0.232\textwidth]{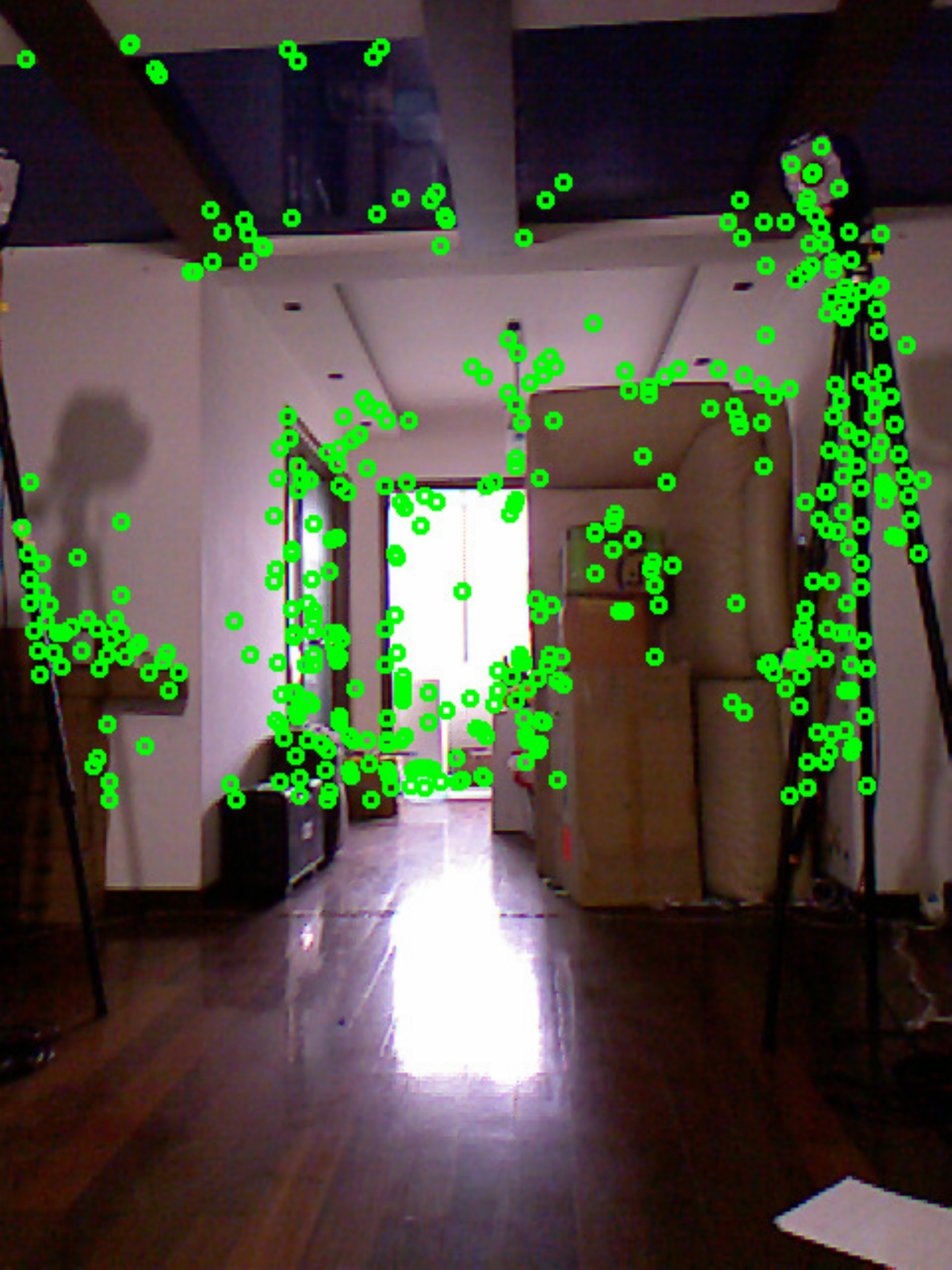}}
		\subfigure[]{\label{fig22} \includegraphics[width=0.232\textwidth]{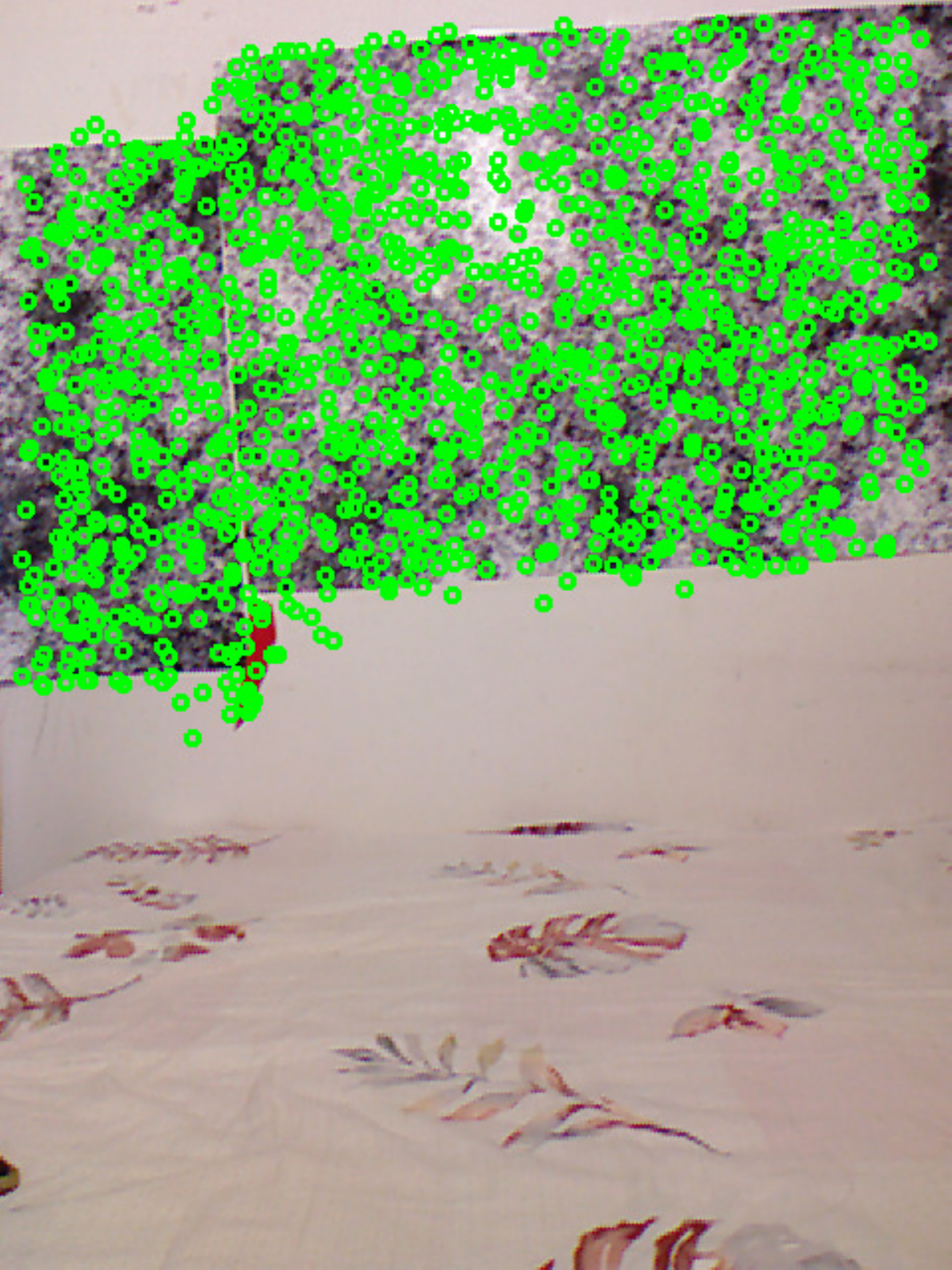}}
		\caption{(a) and (b) show the detected keypoints of the image captured in a natural scene, and the image captured in a scene with descriptor-based patterns respectively. Keypoints are represented by green dots.}
		\label{fig2}
	\end{figure}

	In Fig. \ref{fig2}, the detected SURF keypoints are represented by green dots. Fig. \ref{fig21} and Fig. \ref{fig22} show the detected keypoints of the image captured in a natural scene, and the image captured in a scene with descriptor-based patterns, respectively. There are only only 406 keypoints in Fig. \ref{fig21}, but 1231 keypoints in Fig. \ref{fig22}. 
	
\begin{figure}[t]
		\centering
		\subfigure[]{\label{fig31} \includegraphics[width=0.3\textwidth]{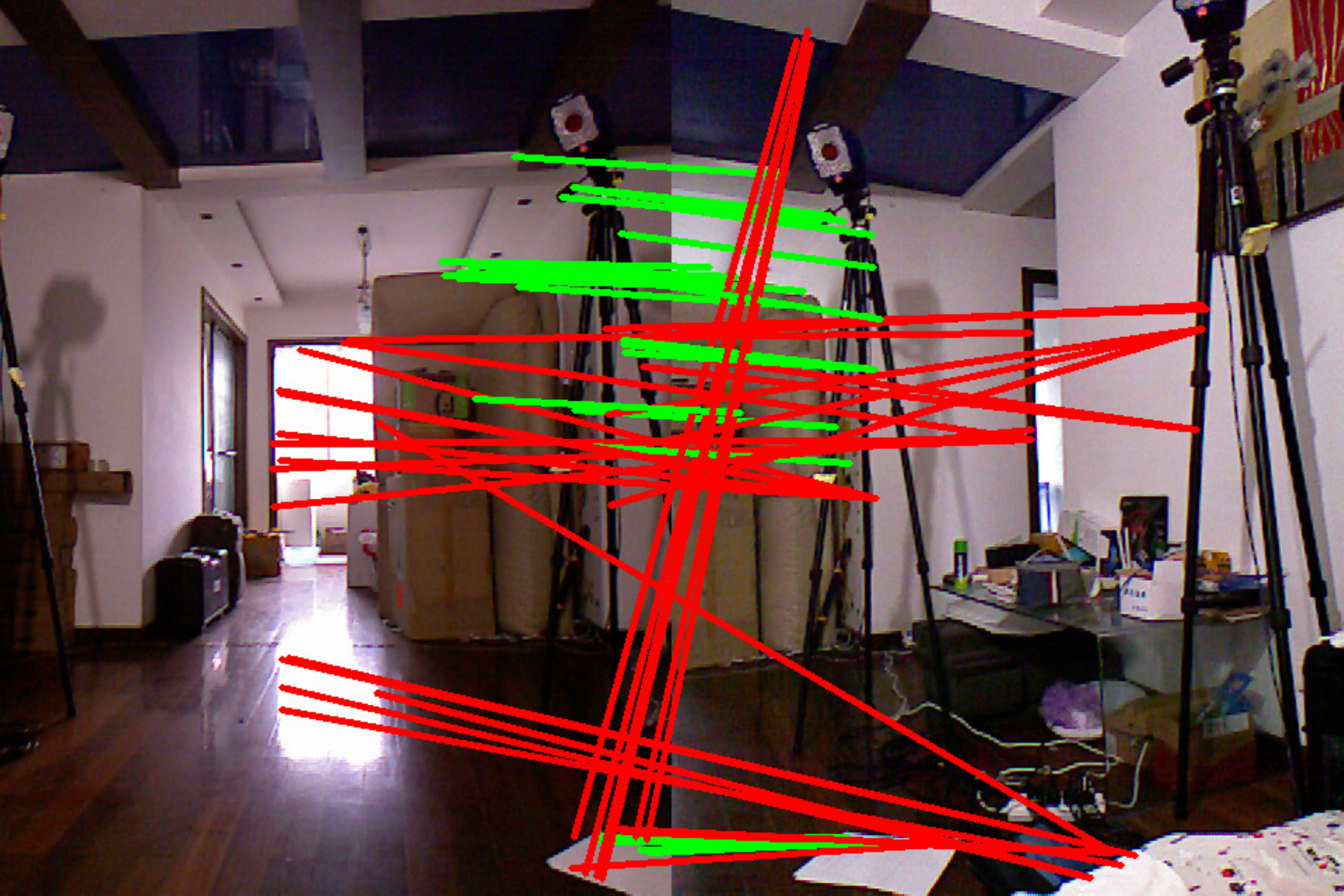}}
		\subfigure[]{\label{fig32} \includegraphics[width=0.3\textwidth]{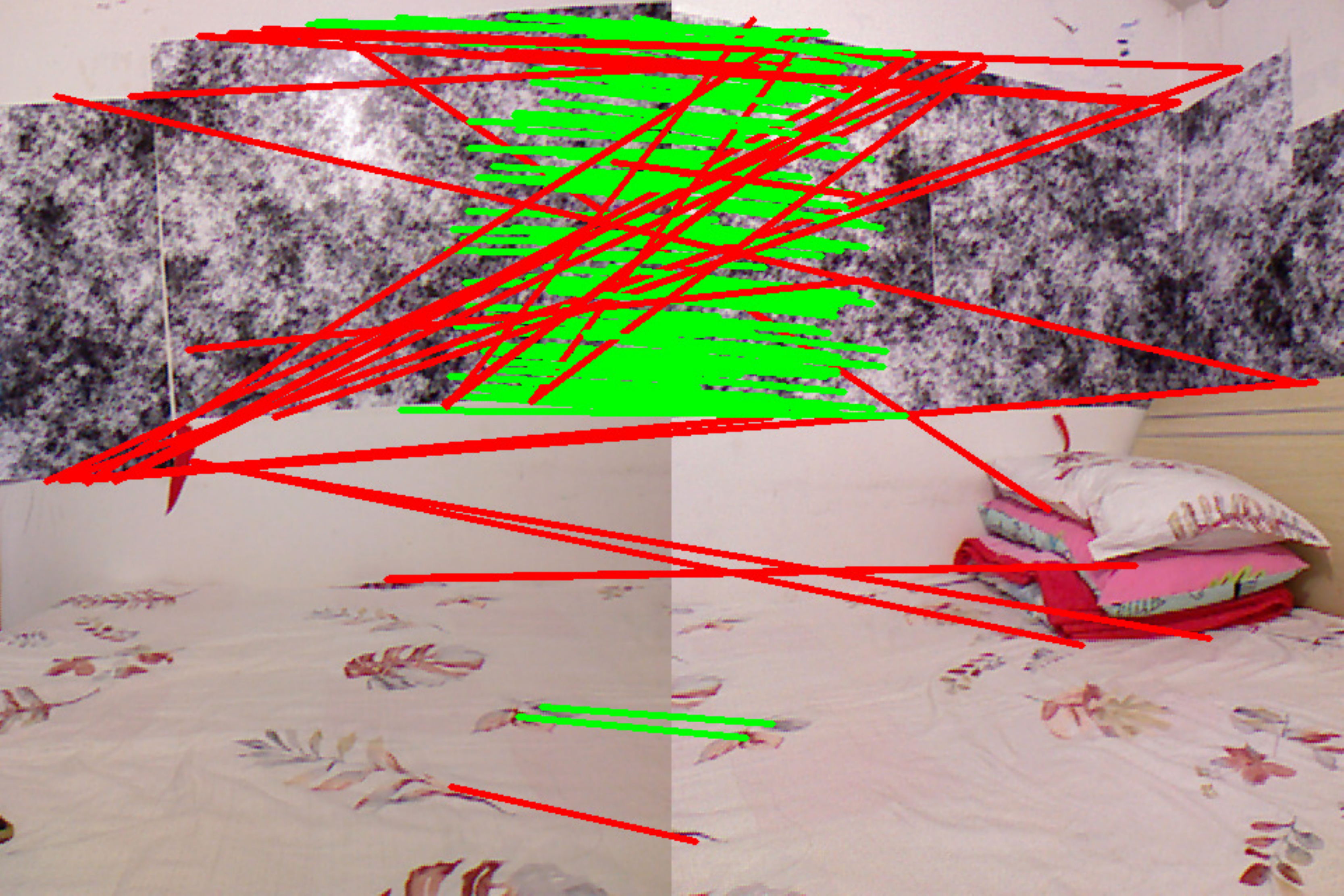}}
		\subfigure[]{\label{fig33} \includegraphics[width=0.3\textwidth]{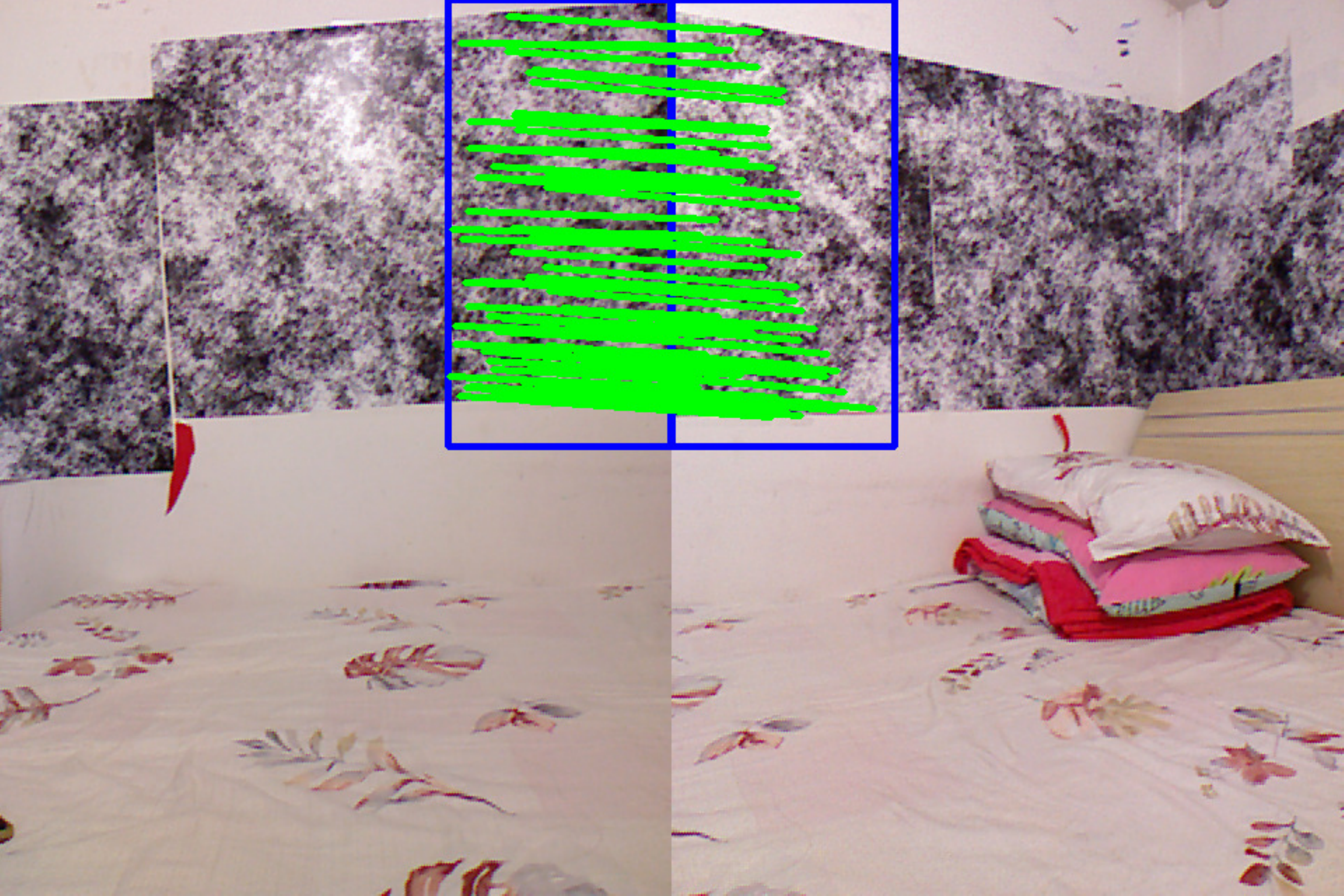}}
		\caption{(a) shows the matched keypoints of the image pairs captured in a natural scene. (b) and (c) show the matched keypoints of the image pairs captured in a scene with descriptor-based patterns. The keypoints in (c) are only detected in the image regions within blue rectangles. Well-matched keypoints are connected with green lines, poorly matched keypoints are connected with red lines.  	
		}
	\end{figure}

	Fig. \ref{fig31} and Fig. \ref{fig32} show the matched keypoints of a pair of images captured by two cameras with an approximately thirty percent overlapping FoV. Well-matched keypoints are connected by green lines; poorly matched keypoints are connected by red lines. The matching results are generated by the Flann-based descriptor matcher implemented in OpenCV, where matched keypoints are accepted only if its descriptor distance is less than 3.0 times the minimum descriptor distances. There are 56 well-matched keypoints and 32 poorly matched keypoints in Fig. \ref{fig31}, 116 well-matched keypoints and 30 poorly matched keypoints in Fig. \ref{fig32}. Thus the descriptor-based patterns provide more robust matched keypoints. In this work, the keypoints detector is only applied to the overlapping regions of neighboring images (see the regions within the blue rectangle in Fig. \ref{fig33}) to reduce the possibility of mismatch and the time of extracting and matching keypoints. From Fig. \ref{fig33}, we can see that all poorly matched keypoints are filtered out.
	
	Based on these matched keypoints, we find their corresponding depth values from the depth maps generated by the depth camera in the RGB-D camera to construct 3D point sets to estimate poses. Due to the different spatial positions and intrinsic parameters of the depth camera and of the color camera in a RGB-D camera, the depth map is not aligned with the color image. The extrinsic parameters between the depth and color camera are used to align the depth map with the color image.
	
	Let $(u_{0}, v_{0})$ denote the coordinates of the principal point of the depth camera, $f_x$ and $f_y$ denote the scale factors in image $u$ and $v$ axes of the depth camera, and $u_{0}$, $v_{0}$, $f_x$ and $f_y$ be the intrinsic parameters of the depth camera. Let $[u,v,Z]$ represent a pixel in the depth map, $Z$ represent the depth value in $[u,v]$, and $[X,Y,Z]^T$ represent the mapped 3D point of $[u,v]$ in the depth camera coordinate system. According to the pinhole camera model, the values of $X$ and $Y$ can be calculated according to
	\begin{equation}
	\begin{array}{l}
	X{\rm{ = }}(u - {u_0})Z/{f_x},\\
	Y = (v - {v_0})Z/{f_y}.
	\end{array}
	\end{equation}
	Let $T$ represent the transformation matrix from the depth camera coordinate frame to the color camera coordinate frame. The relationship between the transformed 3D point ${{\left[ {X}',{Y}',{Z}' \right]}^{T}}$ in the color camera's coordinate frame and ${{\left[ {X},{Y},{Z} \right]}^{T}}$ can be expressed as
	\begin{equation}
	{\left[ {X',Y',Z'} \right]^T} = T{\left[ {X,Y,Z} \right]^T}.
	\end{equation}
	Let $(u_{0}', v_{0}')$ denote the coordinates of the principal point of the color camera and ${f_x}'$ and ${f_y}'$ denote the scale factors in image $u'$ and $v'$ axes of the color camera. After mapping ${{\left[ {X},{Y},{Z} \right]}^{T}}$ to the color image coordinate system, the aligned depth point $[u',v',Z']$ can be obtained, where $u'$ and $v'$ are calculated according to
	\begin{equation}
	\begin{array}{l}
	u' = \frac{{X'}}{{Z'}}{f_x}^\prime  + {u_0}^\prime \\
	v' = \frac{{Y'}}{{Z'}}{f_y}^\prime  + {v_0}^\prime.
	\end{array}
	\end{equation}
	
	We obtain two 3D point sets $\{p_i\}$, $\{p'_i\}$; $i = 1,2,...,N$, based on the matched keypoints and corresponding depth value in the aligned depth map. $p_i$ and $p'_i$ are $3 \times 1$ column matrices. The relative poses between these two 3D point sets can be found by minimizing
	\begin{equation}
\mathop {\min }\limits_\xi   = \frac{1}{2}\sum\limits_{i = 1}^n {\left\| {\left( {{{p'}_i} - \exp (\xi \hat \;){p_i}} \right)} \right\|} _2^2,
	\end{equation}
	where $\xi  \in se(3)$ is a vector with six dimensions that represents the camera pose, $\hat{~}$ maps $\xi$ to a matrix with four mentions ${R^{4 \times 4}}$. $\xi\hat{~}$ is mapped to $T \in SE(3)$ by the exponential map
	$exp()$. We use bundle adjustment to jointly solve all camera poses \cite{kummerle2011g}. The derivative with respect to the camera pose of an error element is given by

	\begin{equation}
	\begin{array}{l}
	\frac{{\partial (\exp (\xi\hat{~} )p)}}{{\partial \xi }} = \mathop {\lim }\limits_{\delta \xi  \to 0} \frac{{\exp (\delta \xi\hat{~} )\exp (\xi\hat{~} )p - \exp (\xi\hat{~} )p}}{{\delta \xi }}\\
	{\kern 1pt} {\kern 1pt} {\kern 1pt} {\kern 1pt} {\kern 1pt} {\kern 1pt} {\kern 1pt} {\kern 1pt} {\kern 1pt} {\kern 1pt} {\kern 1pt} {\kern 1pt} {\kern 1pt} {\kern 1pt} {\kern 1pt} {\kern 1pt} {\kern 1pt}  \approx \mathop {\lim }\limits_{\delta \xi  \to 0} \frac{{(I + \delta \xi\hat{~} )\exp (\xi\hat{~} )p - \exp (\xi\hat{~} )p}}{{\delta \xi }}\\
	{\kern 1pt} {\kern 1pt} {\kern 1pt} {\kern 1pt} {\kern 1pt} {\kern 1pt} {\kern 1pt} {\kern 1pt} {\kern 1pt} {\kern 1pt} {\kern 1pt} {\kern 1pt} {\kern 1pt} {\kern 1pt} {\kern 1pt} {\kern 1pt} {\kern 1pt}  = \mathop {\lim }\limits_{\delta \xi  \to 0} \frac{{\delta \xi\hat{~} \exp (\xi\hat{~} )p}}{{\delta \xi }} = \mathop {\lim }\limits_{\delta \xi  \to 0} \frac{{ - (\exp (\xi\hat{~} )p)\hat{~}\delta \xi }}{{\delta \xi }} \\
	{\kern 1pt} {\kern 1pt} {\kern 1pt} {\kern 1pt} {\kern 1pt} {\kern 1pt} {\kern 1pt} {\kern 1pt} {\kern 1pt} {\kern 1pt} {\kern 1pt} {\kern 1pt} {\kern 1pt} {\kern 1pt} {\kern 1pt} {\kern 1pt} {\kern 1pt}=-(\exp (\xi\hat{~} )p)\hat{~}
	\end{array}
	\end{equation}
where $\hat{~}$ denotes the mapping from ${\exp (\xi\hat{~} )p}$ represented by $[x,y,z]^T$ to the corresponding anti-symmetric matrix
	$\left[ {\begin{array}{*{20}{c}}
		0&{ - z}&y\\
		z&0&{ - x}\\
		{ - y}&x&0
		\end{array}} \right]$.
	
	\subsection{Pose Graph Optimization}
	Assume that there are $N$ RGB-D cameras in the panoramic 3D vision system. Let $T_{i,j}$ denote the estimated relative poses between two adjacent RGB-D cameras, $i = 1,2,...,N$; $j=i+1$, when $i$ equals $N$, $j$ is 1. When the pose of the first RGB-D camera is set to $[0,0,0,1]^T$, we can calculate the pose $x_i$ for each RGB-D camera based on $T_{i,j}$. Due to the pose estimation error, the pose that is calculated by ${T_{N,1}}{x_N}$ is not equal to $[0,0,0,1]^T$. Fortunately, this panoramic setup of RGB-D cameras provides a definite loop-closure constraint for optimizing the estimated poses. This problem can be solved by using the popular pose graph optimization method in SLAM \cite{kummerle2011g, konolige2010sparse}.
	
	According to pose graph optimization theory, the problem can be solved by finding the minimum of a function of this form:
	\begin{equation}
		{x^*} = \arg {\kern 1pt} {\kern 1pt} \mathop {\min }\limits_x \sum\limits_{(i,j) \in S} {r_{i,j}^T} (x){\Lambda _{i,j}}r{}_{i,j}(x),
	\end{equation}
	where $x = {\left( {x_1^T,{\kern 1pt} ...,x_n^T} \right)^T}$ is a vector of poses, $r{}_{i,j}$ is the residual of the predicted and observed relative poses between the $i-$th and $j-$th node, ${\Lambda _{i,j}}$ denotes the measurement information matrix, and $S$ represents the set of edges that connect the nodes.
	
	\section{EXPERIMENTAL RESULTS}
	\label{sec3}
	Experiments were performed to evaluate the accuracy of the proposed extrinsic calibration method and demonstrate the efficiency of constructing indoor environments using our panoramic RGB-D camera setup. We designed a camera rig that consists of twelve Kinect v1 RGBD cameras mounted in a radial configuration (see Fig. 1(a)).
	
\begin{figure}[t]
		\centering
		\includegraphics[width=0.35\textwidth]{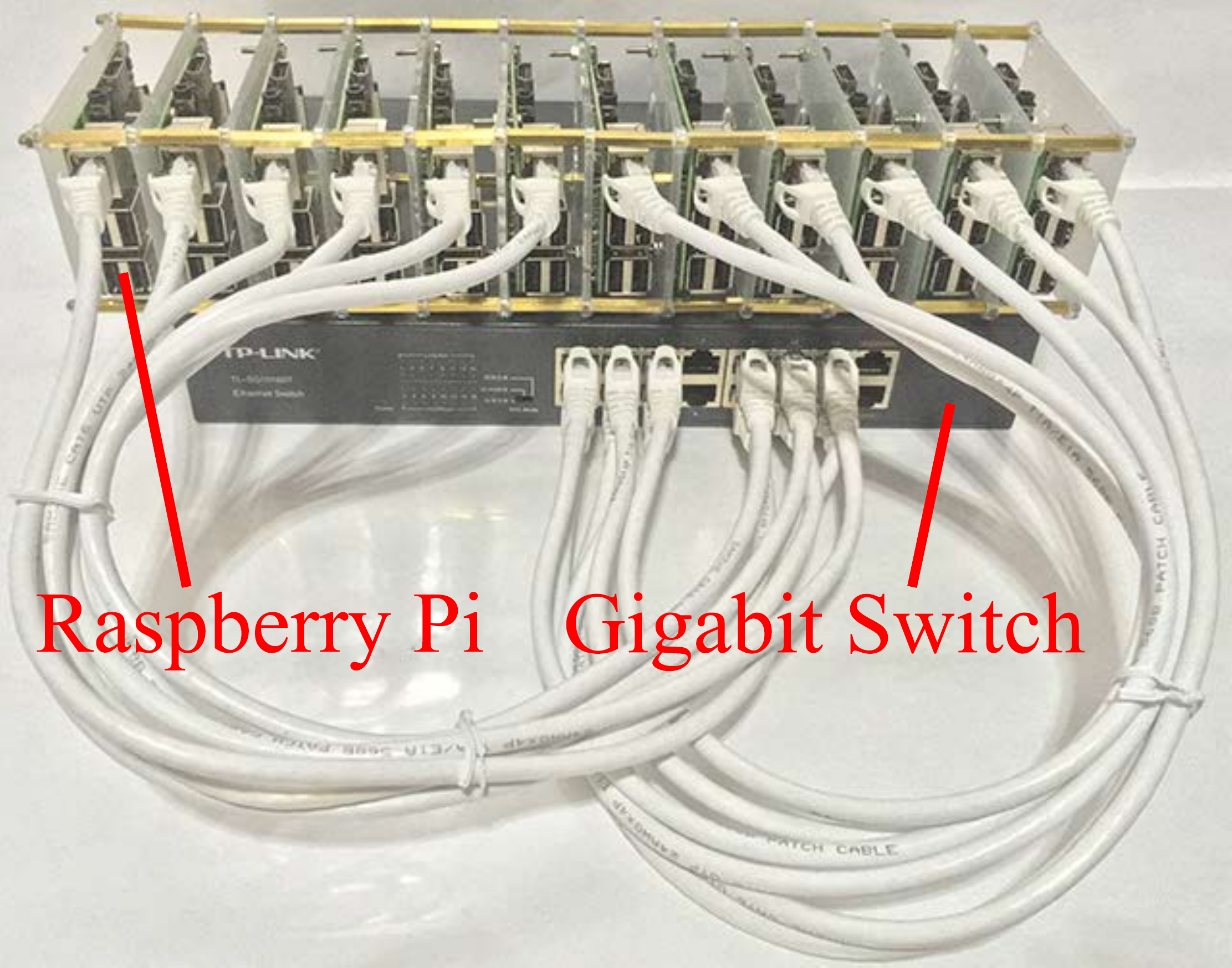}
		\caption{Distributed capturing system consisting of a gigabit switch and 12 low-cost Raspberry Pi single board computers.}
		\label{fig3}
	\end{figure}

	We also designed a distributed system built on top of a local area network (LAN) to capture the RGB-D frames from all twelve cameras in real time. The distributed capturing system (see Fig. \ref{fig3}) consists of twelve Raspberry Pi single board computers, a gigabit switch and a PC. Raspberry Pi is used to obtain RGB-D frames from each Kinect and send the data to the PC through the LAN using User Datagram Protocol. The PC is used to receive and process the RGB-D frames. Both the depth and color frames were set to a size of $640 \times 480$; the values of a single pixel in the depth frame and color frame have a size of 2 bytes and 3 bytes, respectively. Thus, the size of the RGB-D frames from all twelve Kinects is 17.58 million bytes, which can be sent to the PC through the gigabit switch at 7 fps. This frame rate can be increased by decreasing the size of RGB-D frames from Kinect.
	
	\begin{figure}[t]
		\centering
		\subfigure[]{\includegraphics[width=0.2695\textwidth]{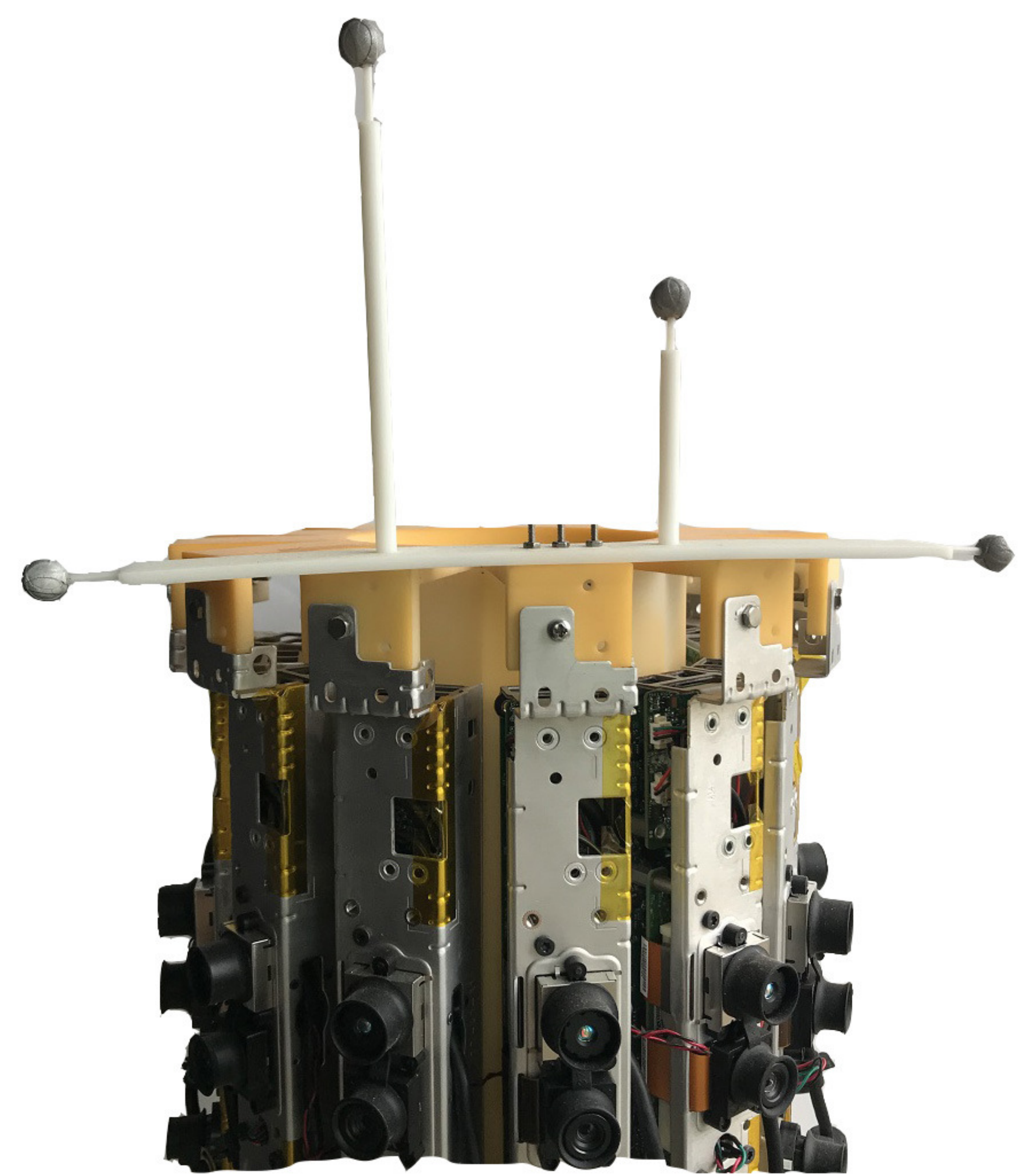}} \\ 
		\subfigure[]{\includegraphics[width=0.2695\textwidth]{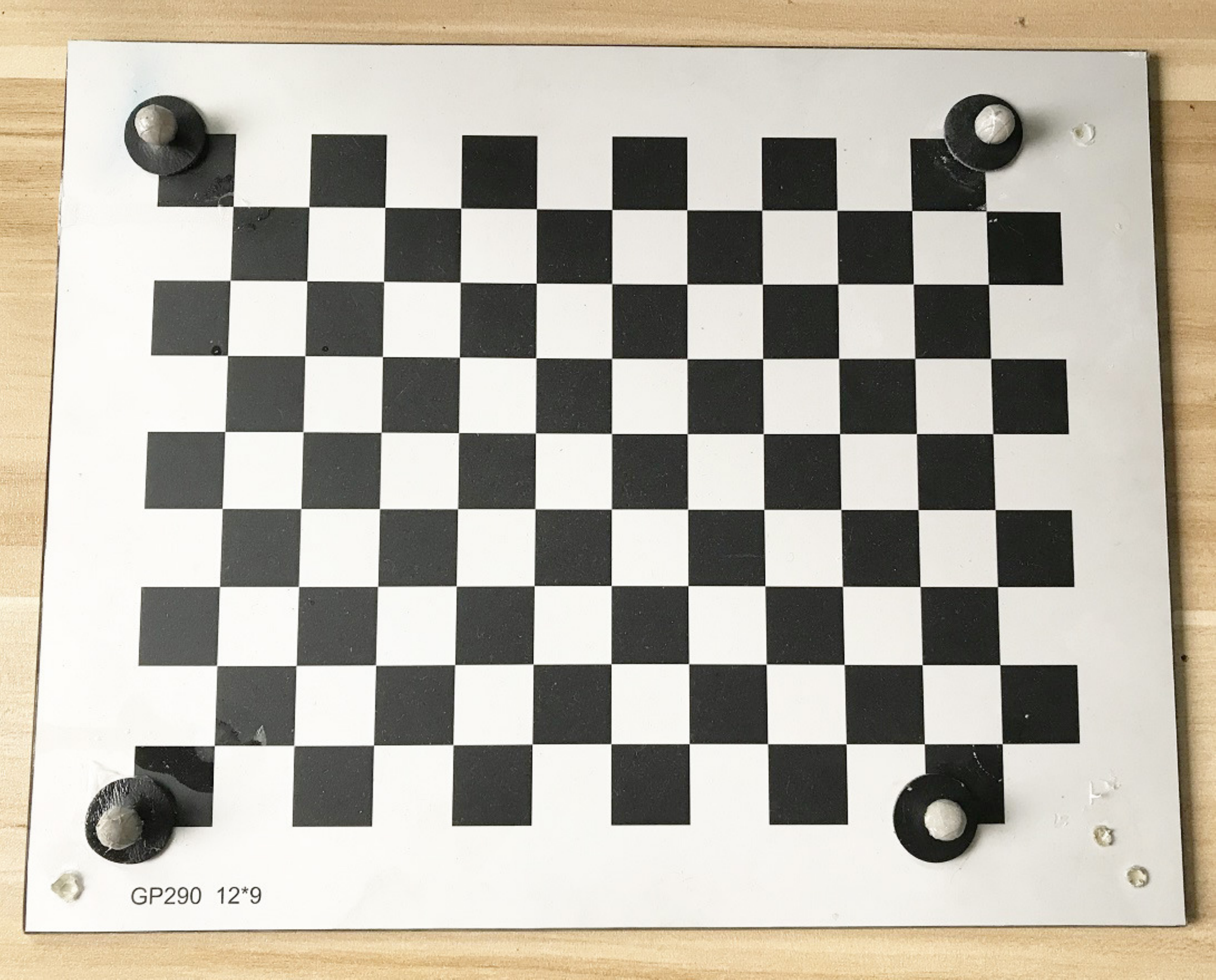}}
		\caption{(a) Kinect with reflective markers. (b) Chessboard with reflective markers.}
		\label{fig4}
	\end{figure}
	
	To evaluate the accuracy of the proposed extrinsic calibration method, we used a motion capture system to obtain the ground truth relative poses between the cameras. The motion capture system requires at least three reflective markers to track the pose of a rigid body, such as the Kinect and the chessboard in our experiments. We attached four reflective markers to both the Kinect and the chessboard (see Fig. \ref{fig4}). We placed the four markers on the outer corners of the chessboard such that the relative poses between the chessboard and the motion capture system and the Kinect color camera were known. The motion capture system tracked the poses of the markers that were attached to the Kinect to determine the poses of the Kinect color camera. For convenience, we only attached markers to one Kinect and placed this Kinect at twelve different positions in the camera rig to capture RGD frames. The accuracy of extrinsic calibration was evaluated using twelve pairs of RGB-D frames from the captured twelve RGB-D frames.
		
	\begin{figure*}[thbp]
		\centering
		\subfigure[]{\label{fig51}\includegraphics[width=0.3\textwidth]{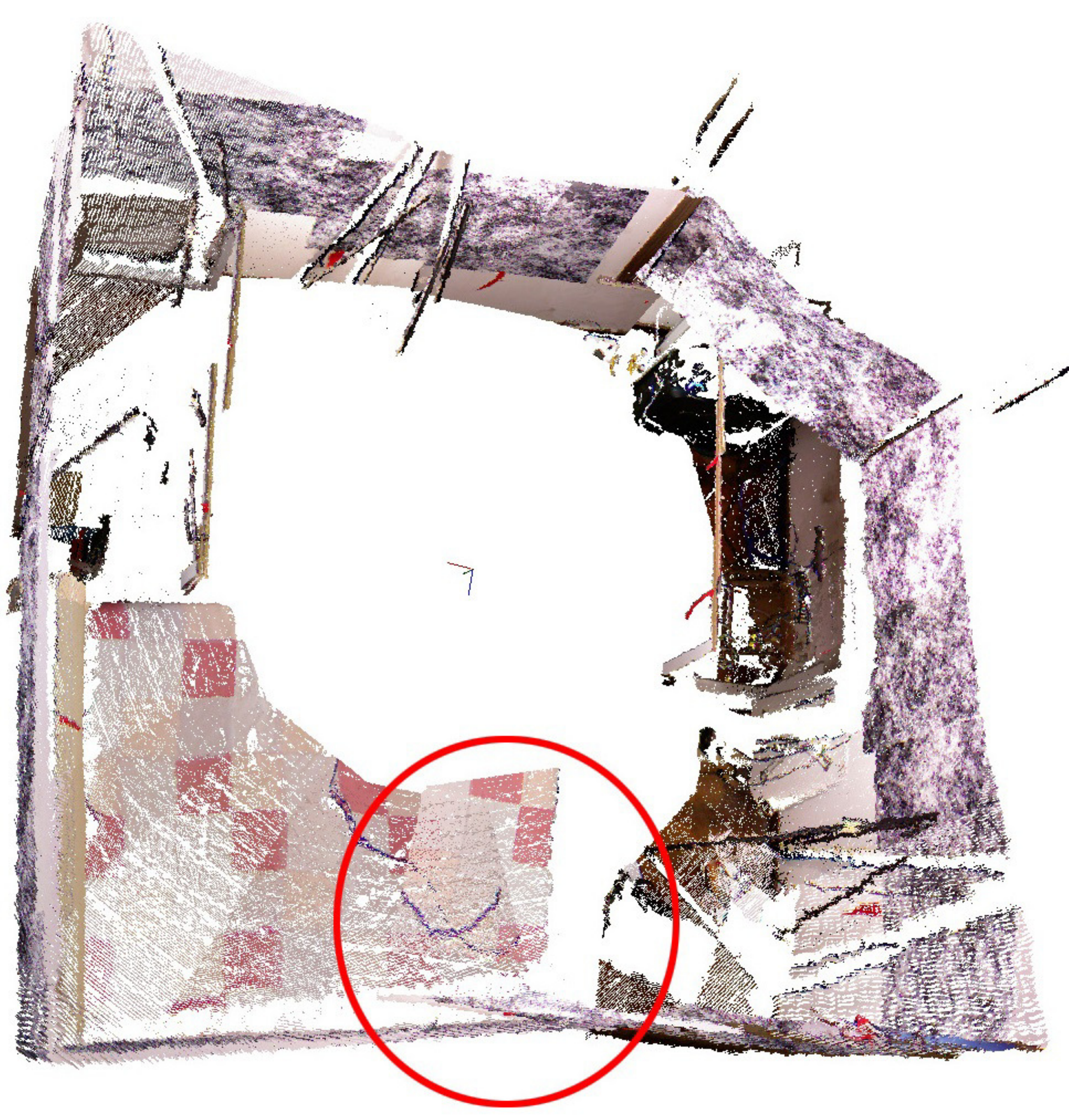}}
		\subfigure[]{\label{fig52}\includegraphics[width=0.3\textwidth]{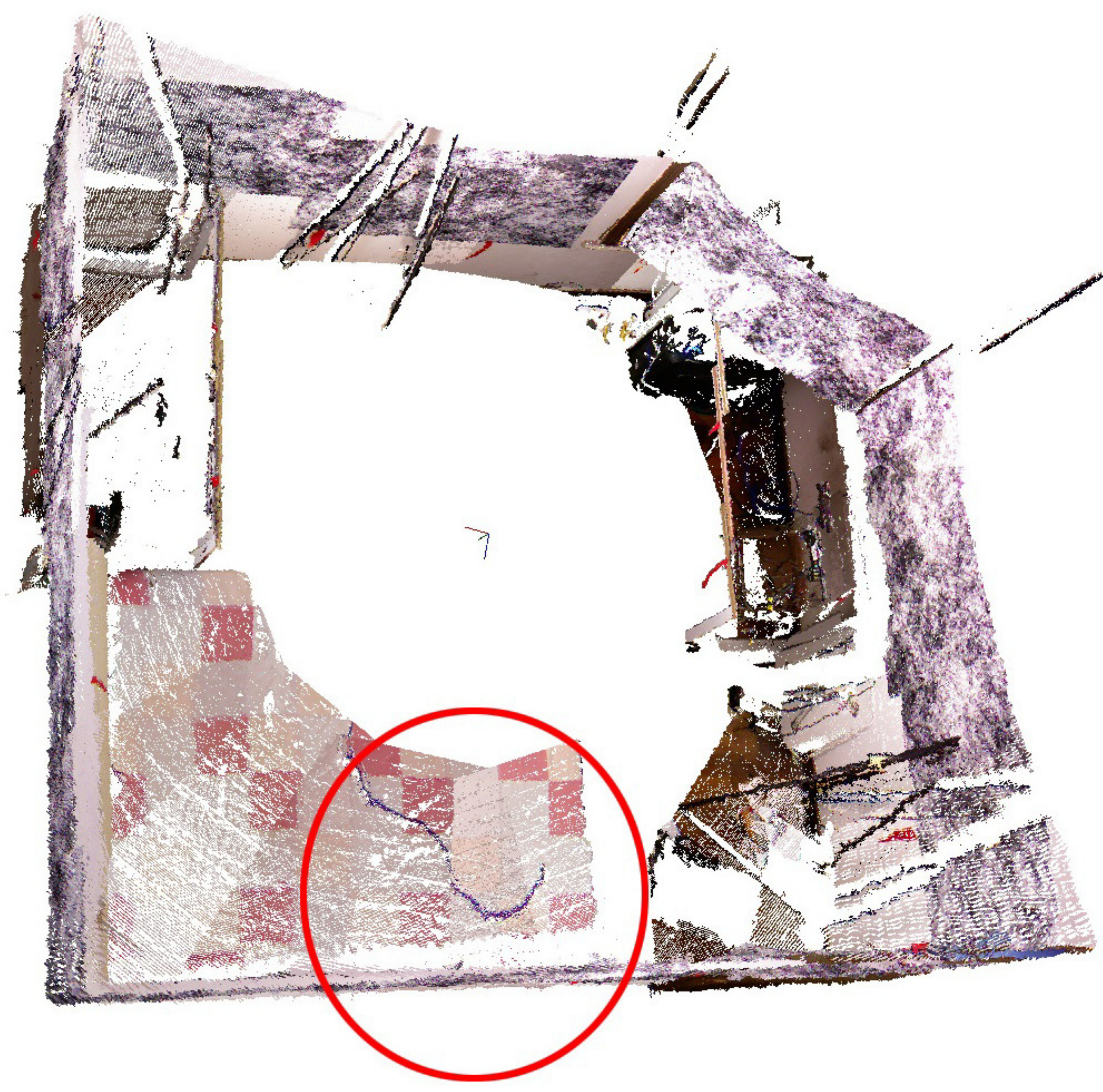}}
		\caption{The aligned point clouds before (a) and after (b) closing the loop. A misalignment can be observed in the red circle on the bottom of (a). After pose graph optimization with loop closure, the misalignment is resolved.}
		\label{fig5}
	\end{figure*}
		
	\begin{table*}[thpb]
		\caption{Residual errors of initially estimated poses and optimized poses}
		\begin{center}
			\begin{tabular}{|c|c|c|c|c|c|c|c|c|c|c|}
				\hline
				$dist\_thresh$/$min\_dist$ & 1.5 & 2 & 3 & 4 & 5 & 6 & 7 & 8 & 9 & 10 \\
				\hline
				Correspondences & 14 & 56 & 218 & 373 & 477 & 568 & 615 & 631 & 642 & 651 \\
				\hline
				Ini. rot error (deg) & 2.07&1.48&1.40&1.48&1.51&1.53&1.51&1.52&1.52&1.52 \\
				\hline
				Opti. rot error (deg)&1.96&0.89&0.76&0.62&0.61&0.56&0.56&0.56&0.56&0.56 \\
				\hline
				Ini. trans error (cm)&9.25&2.73&1.86&2.07&2.07&1.81&1.80&1.81&1.80&1.81 \\
				\hline
				Opti. trans error (cm)&4.94&2.78&1.88&1.84&1.77&1.83&1.83&1.81&1.82&1.80 \\
				\hline
			\end{tabular}
		\end{center}
		\label{table_one}
	\end{table*}

	\begin{figure*}[thpb]
		\centering
		\subfigure[]{\label{fig61}\includegraphics[width=0.4\textwidth]{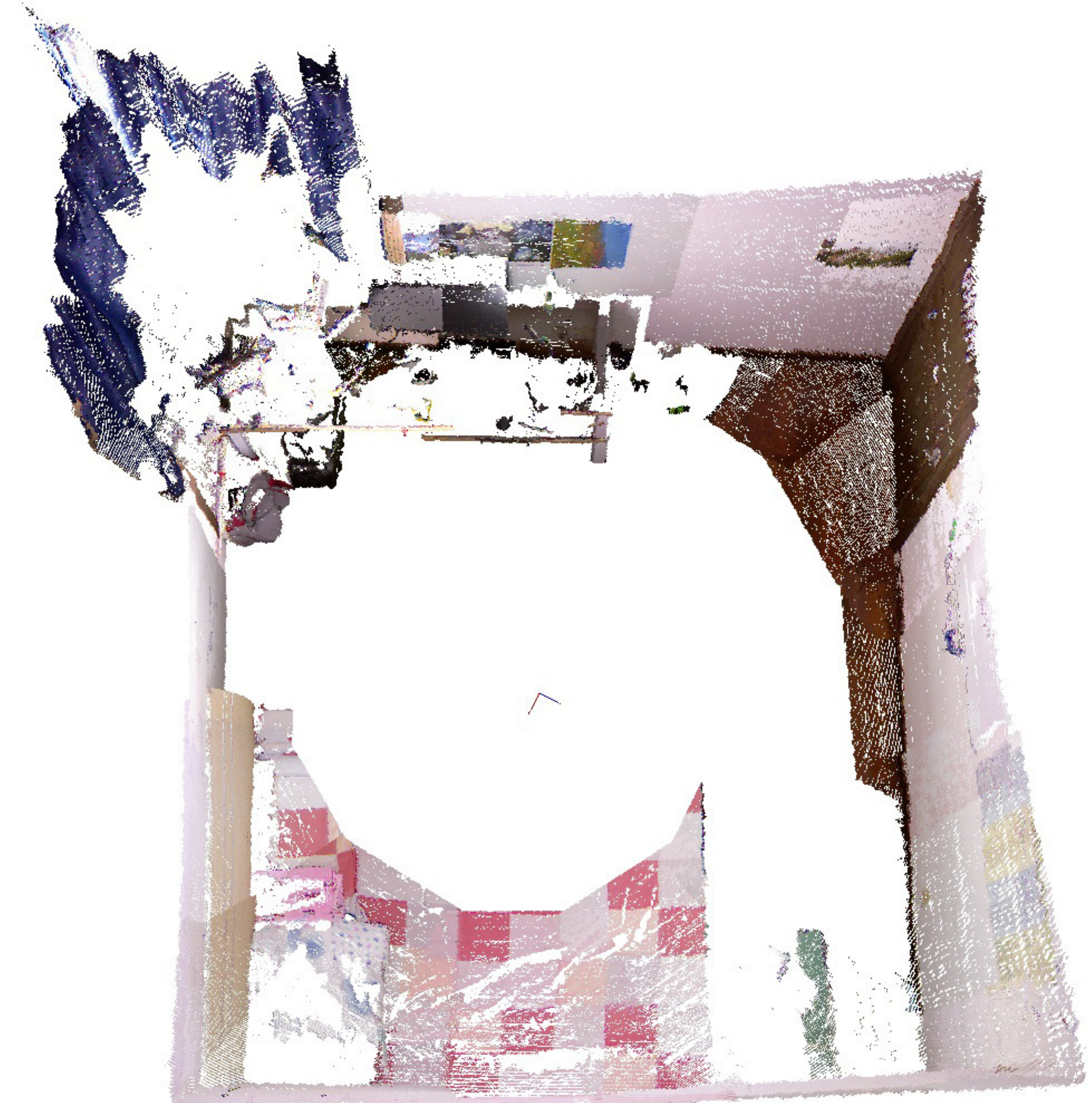}} 
		\subfigure[]{\label{fig62}\includegraphics[width=0.4\textwidth]{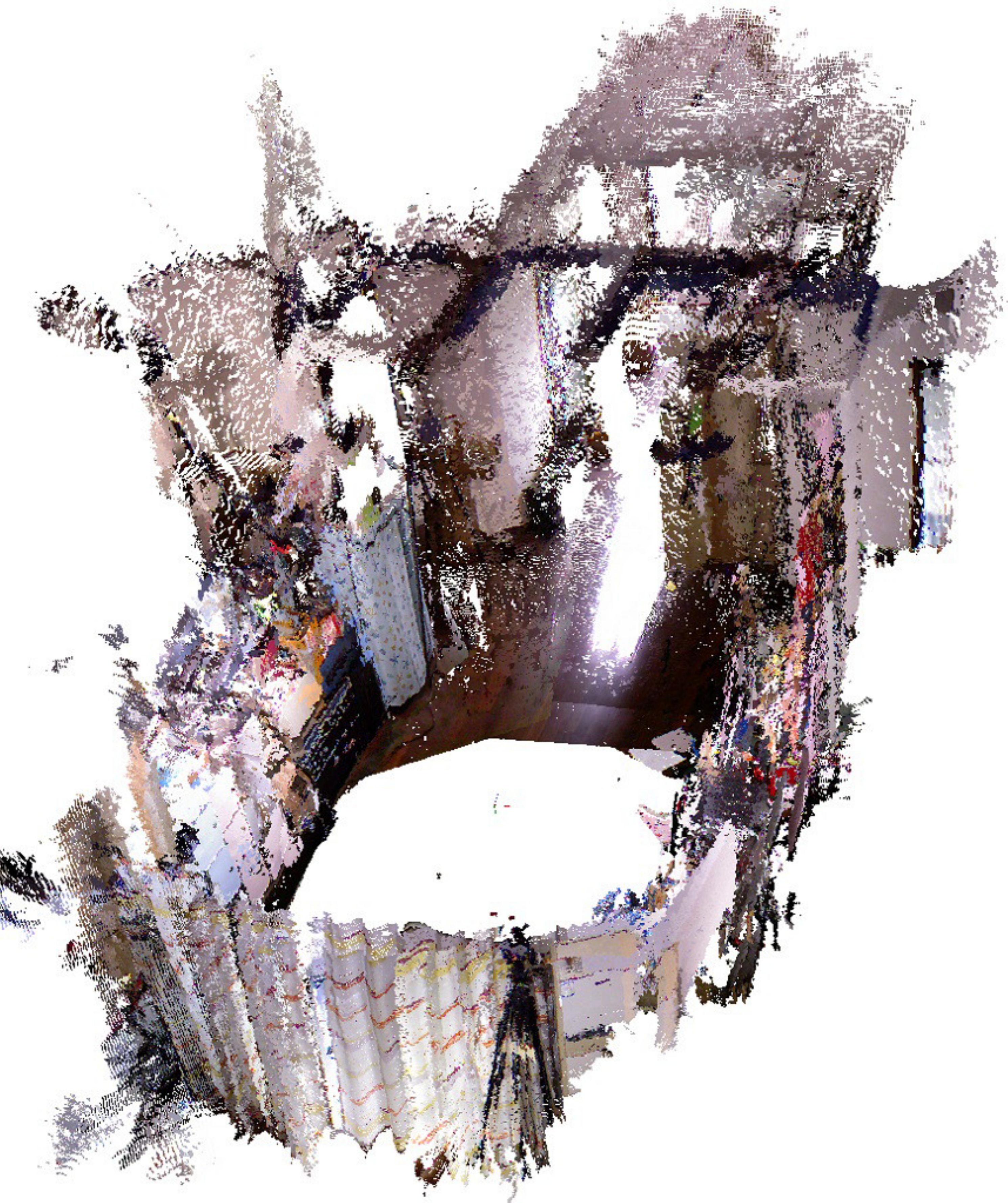}}
		\caption{Reconstruction of a bedroom (3.3 m $\times$ 3 m) (a), and a living room (9 m $\times$ 3.5 m) using only one output frame from the proposed panoramic 3D vision system.}
		\label{fig6}
	\end{figure*}

	While determining the correspondences of keypoints for each pair of RGB-D frames, we increased the distance threshold of the keypoints (denoted by $dist\_thresh$) from one and a half times the minimum distance to the maximum distance (ten times the minimum distance on average) to analyze the calibration error in rotation and translation with respect to the keypoints correspondences. In TABLE \ref{table_one}, the average residual error in rotation and translation is presented for the initial estimation (denoted by Ini.) and optimized estimation (denoted by Opti.) using pose graph optimization with loop closure. It can be seen that the average residual error was reduced when raising the distance threshold to consider more keypoint correspondences. The residual error in both rotation and translation for the optimized estimation is generally less than the residual error for the initial estimation. When the distance threshold was set to be larger than four times the minimum distance, the residual error remained at the same level. The aligned point clouds before and after pose graph optimization with loop closure can also be observed in Fig. \ref{fig5}. It can be seen in the red circle on the bottom of Fig. \ref{fig51} that the point clouds were misaligned; however, these point clouds aligned well in Fig. \ref{fig52} with optimized poses with the loop closure constraint.
	 
	The intrinsic calibration process can be avoided as the intrinsic parameters of Kinects can be obtained through the software development kit (SDK) of Kinect. The SDK also provides an application interface to map the depth map to color camera coordinate frame in real time. The whole calibration process using our setup and method cost 800 ms on a computer with a 3.60 GHz CPU, most of which was spent on extracting and matching feature points. In comparison, for the setup and method reported in \cite{Fernandez2014Extrinsic}, the panoramic RGB-D camera setup was required to be moved around to take about 200 images in order to extract enough planes for all camera pairs to estimated poses, which took more than 5 seconds \cite{Fernandez2014Extrinsic}. In terms of accuracy, the residual error of rotation was 1.60 degrees, and the error of translation was 2.5 cm in \cite{Fernandez2014Extrinsic}. Our proposed calibration method resulted in a rotation error of 0.56 degree and a translation error of 1.80 cm (see TABLE \ref{table_one}).
	
	The operating range of the Kinect v1 depth camera is between 0.5 m to 5.0 m \cite{Khoshelham2012Accuracy}. Using an output panoramic frame can reconstruct scenes within the circle with a radius of five meters. Thus the reconstruction of indoor scenes becomes very efficient using our panoramic RGB-D camera setup. Fig. \ref{fig6} presents the reconstruction result of a bedroom (3.3 m $\times$ 3 m) and a living room (9 m $\times$ 3.5 m) using only one output panoramic frame from the system. The panoramic RGB-D camera setup provides a 360$^\circ$ FoV, leading to better constraints for localization and can be potentially used to reduce localization and mapping errors. Our next step is to investigate direct registration methods such as \cite{engel2017direct, forster2014svo}, which do not depend on time-consuming keypoint detectors or descriptors for large scale SLAM, using this panoramic 3D vision system.
	
	\section{CONCLUSION}
	
	In this letter, a new method that relies on well-matched keypoints provided by a feature descriptor-based calibration pattern was proposed to calibrate the extrinsic parameters of the RGB-D cameras in the system. A LAN-based distributed system was developed, which enabled the system to provide panoramic RGB-D frames in real time. The reconstruction of indoor scenes was efficiently and conveniently performed using the panoramic RGB-D 3D vision system. The experiments validated the accuracy and efficiency of the proposed calibration method and the efficiency of the panoramic RGB-D camera setup in 3D reconstruction, and quantitatively demonstrated a higher speed and higher accuracy compared with existing methods.

	\section{ACKNOWLEDGMENTS}
	We thank Professor Min Li and Zhihong Huang for assistance in constructing and testing the motion capture system, Dr. Xiang Gao et al. for the easily understandable explanation about SLAM in \cite{Gao2017SLAM}. This work was supported by the National Natural Science Foundation of China (grant numbers 61525305 and 61625304), the Shanghai Natural Science Foundation (grant numbers 17ZR1409700 and 18ZR1415300), and the basic research project of Shanghai Municipal Science and Technology Commission (grant number 16JC1400900).
	
	\bibliographystyle{bibliography/IEEEtran}
	\bibliography{bibliography/IEEEabrv,bibliography/mybibfile}\ 
	

\end{document}